\documentclass{elsart5p}

\usepackage{natbib}

 \usepackage{graphicx}

\usepackage{amssymb}

\begin{document}

\begin{frontmatter}



\title{Cosmic ray composition at high energies: The TRACER project}
\author{P.J. Boyle, for the TRACER project\thanksref{aa}}
\address{Enrico Fermi Institute, University of Chicago, 5640 S Ellis Ave, Chicago, Illinois 60637, USA}
\thanks[aa]{M. Ave, P.J. Boyle, F. Gahbauer, C. H\"{o}ppner, J. H\"{o}randel, M. Ichimura, D. M\"{u}ller, A. Romero Wolf}


\begin{abstract}

The TRACER instrument (\emph{Transition Radiation Array for Cosmic
Energetic Radiation}) is designed to measure the individual energy
spectra of cosmic ray nuclei in long duration balloon flights. The
large geometric factor of TRACER (5 m$^2$ sr) permits statistically
significant measurements at particle energies well beyond
10$^{14}$eV. TRACER identifies individual cosmic ray nuclei with
single element resolution, and measures their energy over a wide
range, from about 0.5 to 10,000 GeV/amu. This is accomplished with
a gas detector system of 1600 single wire proportional tubes and
plastic fiber radiators combined with plastic scintillators and
acrylic Cerenkov counters. A two week flight in Antarctica in December
2003 has led to a measurement of the nuclear species oxygen to iron up
to about 3,000 GeV/amu. As an example, we shall present the energy
spectrum and relative abundance for neon and discuss the implication
of this result in the context of current models of acceleration and
propagation of galactic cosmic rays. The instrument has been
refurbished and flown on a second long duration balloon flight in
Summer 2006. For this flight, the dynamic range of TRACER has been
extended to permit inclusion of the lighter elements B, C and N in the
measurement.

\end{abstract}


\end{frontmatter}


\section{Introduction}
\label{intro}

Many questions still remain about the origin of cosmic radiation even
after nearly 100 years since its discovery. The radiation covers a
very large range in energy, 10$^9$ to 10$^{20}$ eV, but exhibits a
surprisingly smooth powerlaw energy spectrum, with just a modest break
or 'knee' around 10$^{15}$ eV. It had been proposed by \cite{baade34}
that cosmic rays may be generated by supernovae, while Fermi, in his
seminal paper in 1949, put forward a mechanism for acceleration due to
collisions with magnetized clouds. The Fermi mechanism would produce a
power law energy spectrum but could not account for the presence of
heavier elements (Z $>$ 3) in cosmic rays. In the late 1970s, these
basic ideas were reconciled with a model of acceleration in
interstellar shock fronts caused by supernova explosions
(\cite{bell78}). For strong shocks, this model predicts a powerlaw
energy spectrum with a slope close to $E^{-2.0}$. However, there
appears to be an inherent upper limit to the energies attainable by
this mechanism, around Z $\times 10^{14}$ eV (\cite{lagage83}). One
might speculate that this limit is reflected in the so called 'knee'
in the energy spectrum at 10$^{15}$ eV.

Even at $10^{15}$eV, the gyroradius of cosmic rays in the galaxy is
much smaller than the distance to a potential source and so their
arrival direction will not reveal the position of their source. To
obtain astrophysical information, one measures the absolute intensity
and the elemental, and if possible, isotopic composition of cosmic rays
as a function of energy. Below the knee, the energy spectra of cosmic
ray nuclei are much softer (approximately $\sim E^{-2.7}$) than the
expected spectrum from strong shocks. Hence, if the shock acceleration
model is correct the propagation of cosmic rays through the galaxy
must be dependent on energy.

Indeed, it has been known for a long time (\cite{julliusson73a}),
that the relative abundance of secondary nuclei produced by
fragmentation decreases with energy, indicating that the path length of
material encountered by cosmic rays propagating through the
interstellar medium decreases proportional to $E^{-0.6}$. Detailed
analysis of previous composition experiments (\cite{engelmann90},
\cite{muller91}) has inferred a source energy spectrum $\propto
E^{-2.2}$ (\cite{swordy93}), which is roughly consistent with the
shock acceleration model. However, the energy dependence of the cosmic
ray pathlength is supported by measurements only up to about
$10^{12}$eV/particle. Hence, detailed measurements at higher energies
are needed.

\begin{figure}
\includegraphics[width=.53\textwidth]{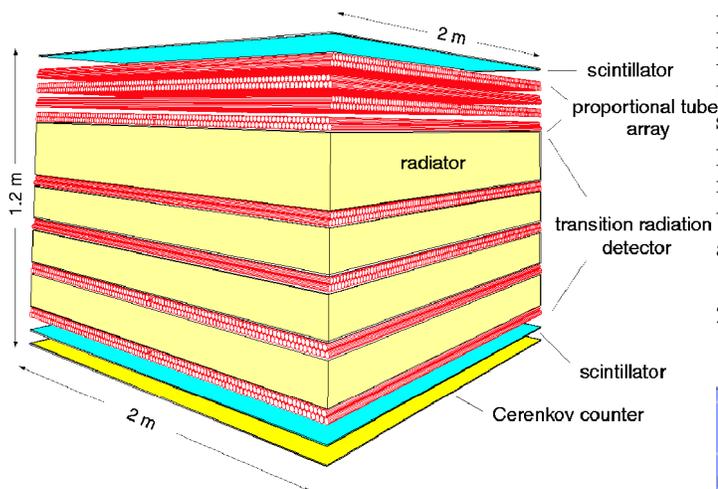}
\caption{Schematic drawing of TRACER 2003}
\label{tracer}
\end{figure}

\section{Description of the TRACER instrument}
\label{instrument}

Direct measurements of the flux of particles at high energy require the
availability of an instrument with large exposure factor. This
requirement is very difficult to attain with conventional
detectors. However, Transition Radiation Detectors (TRDs) may exhibit
large area at relatively low weight, and may have good energy response
up to particle Lorentz factors of $\gamma \approx 10^5$
(\cite{wakely02}). A TRD was first employed in space in the Chicago
``Cosmic Ray Nuclei'' (CRN) experiment flown aboard the Space Shuttle
in 1985 (\cite{swordy90}). The results from this experiment
(\cite{muller91}) still provide the most detailed information on
cosmic ray composition at high energies. After the Challenger
disaster, continued shuttle flights of CRN were no longer possible,
but long duration balloon flights offered a less costly
alternative. However, CRN required the use of a heavy pressure gondola
to maintain the sensitive multiwire proportional chambers (MWPC) in an
environment at atmospheric pressure. This setup is undesirable for
balloon flights because the weight of the pressure vessel ($\approx$
1,500 Kg) would represent almost half the allowed science weight of a
balloon payload. Therefore we replace the MWPC's with arrays of single
wire proportional tubes (SWPT), which can be operated in external
vacuum.

The TRACER experiment is the first balloon borne cosmic ray
composition experiment developed using SWPTs to detect transition
radiation (\cite{gahbauer04}). A design of spiral wound mylar tubes
2cm in diameter, 2m in length and a thickness of about 127 $\mu$m was
chosen (\cite{gahbauer03a}). Each tube contains a 50 $\mu$m stainless
steel wire stretched through its center. Tubes are arranged into
manifolds, consisting of a double layer of 99 tubes, and are filled
with a gas mixture containing xenon and methane. The instrument
contains 16 manifolds with a total of 1584 tubes (see figure
\ref{tracer}). Manifolds are placed in 8 alternative X-Y layers to
provide tracking information. Four upper layers (800 tubes) are
designed to measure the specific ionization of the incoming cosmic ray
particle only, while the four lower layers are combined with
radiators, consisting of blankets of plastic fibers, to produce
transition radiation. The total stack of proportional tubes and
radiator fibers measures 2m x 2m x 1.2m.  Complementing these counters
are plastic scintillation counters that act as a trigger and also
measure the charge of each particle, and a Cerenkov counter that
provides information on the particle's charge and also an energy
measurement in the region of 1-10 GeV/amu.


\subsection{Antarctic Balloon Flight}

\begin{figure}[h]
\begin{center}
\includegraphics[width=.5\textwidth]{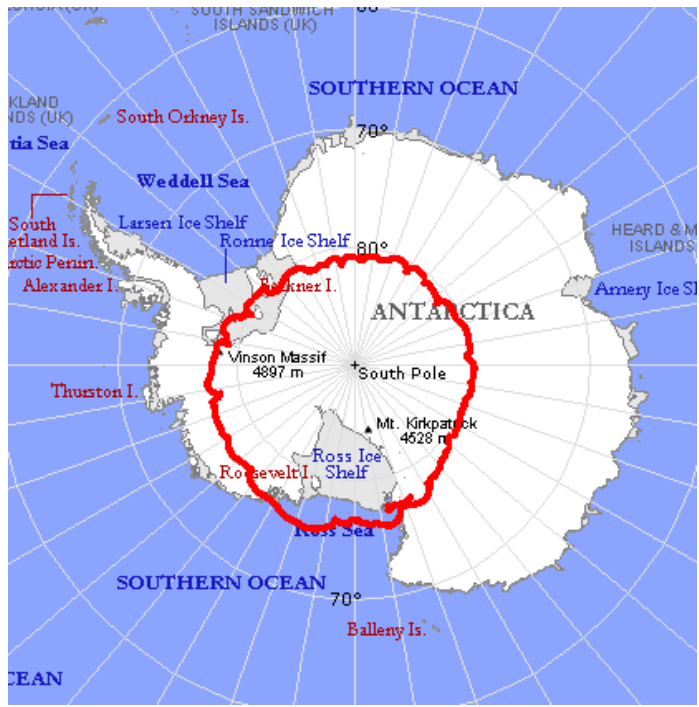}
\end{center}
\caption{Trajectory of 2003 TRACER Antarctic Long Duration Balloon Flight.}
\label{TRdEdx}
\end{figure}

TRACER was exposed above the atmosphere during a long duration balloon
flight of 14 days in Antarctica in December 2003
(\cite{romero05a}). The flight provided a total exposure of 50 m$^2$
steradian days at an average altitude of 125,000 feet or 3.9 g/cm$^2$
residual atmosphere. The instrument performed well during the
flight. There were no problems with high voltage discharge or corona,
and temperature sensors throughout the instrument recorded values in
the range 10-25$^\circ$C, entirely consistent with a thermal analysis
carried out before the flight (\cite{cannon03}). The quality of the
gas in the TRD was monitored during flight and was found to permit
stable gas amplification, with variations less than 3\%. This
performance is much better than that observed on ground, where the
gain of the tubes deteriorates with time constants of the order of a
day, most likely due to diffusion of oxygen into the tubes. The
instrument was fully recovered and returned to Chicago, where a
refurbishment for a second long duration balloon flight was
subsequently carried out and is described in Section 5.

\section{Analysis}

\subsection{Trajectory Reconstruction}

The first step in the analysis of the data is the accurate
reconstruction of the cosmic ray trajectories through the
instrument. As TRACER has no independent tracking devices we use the
entire proportional tube array for this purpose. As a first estimate
the path is obtained using the center of each of the tubes hit in an
event. This allows an accuracy of 5mm in track position. As a second
step, the trajectory is refined by using the fact that the energy
deposit in each tube is proportional to the track length in that
tube. With this method we achieve an accuracy of 2mm in the lateral
track position, corresponding to 3\% in total pathlength through all
the tubes. The accuracy and efficiency of this method are verified
with a GEANT4 simulation of the instrument. Once the trajectory is
known, the magnitude dE/dx is determined :

\vspace{0.2cm}
\begin{center}
\begin{equation}
\frac{dE}{dx} = \frac{\sum_i dE_i}{\sum_i dx_i} = \frac{\sum \mbox{energy deposit in tube(i)}}{\sum \mbox{pathlength in tube(i)}}
\end{equation}
\end{center}
\vspace{0.2cm}

where the summation goes over all tubes. A cut of dx$_i$ $>$ 1 cm is
applied to avoid the large fluctuations associated with short
pathlengths.

\subsection{Charge Analysis}

\begin{figure}[h]
\begin{minipage}[t]{.45\textwidth}
\begin{center}
\includegraphics[width=.99\textwidth]{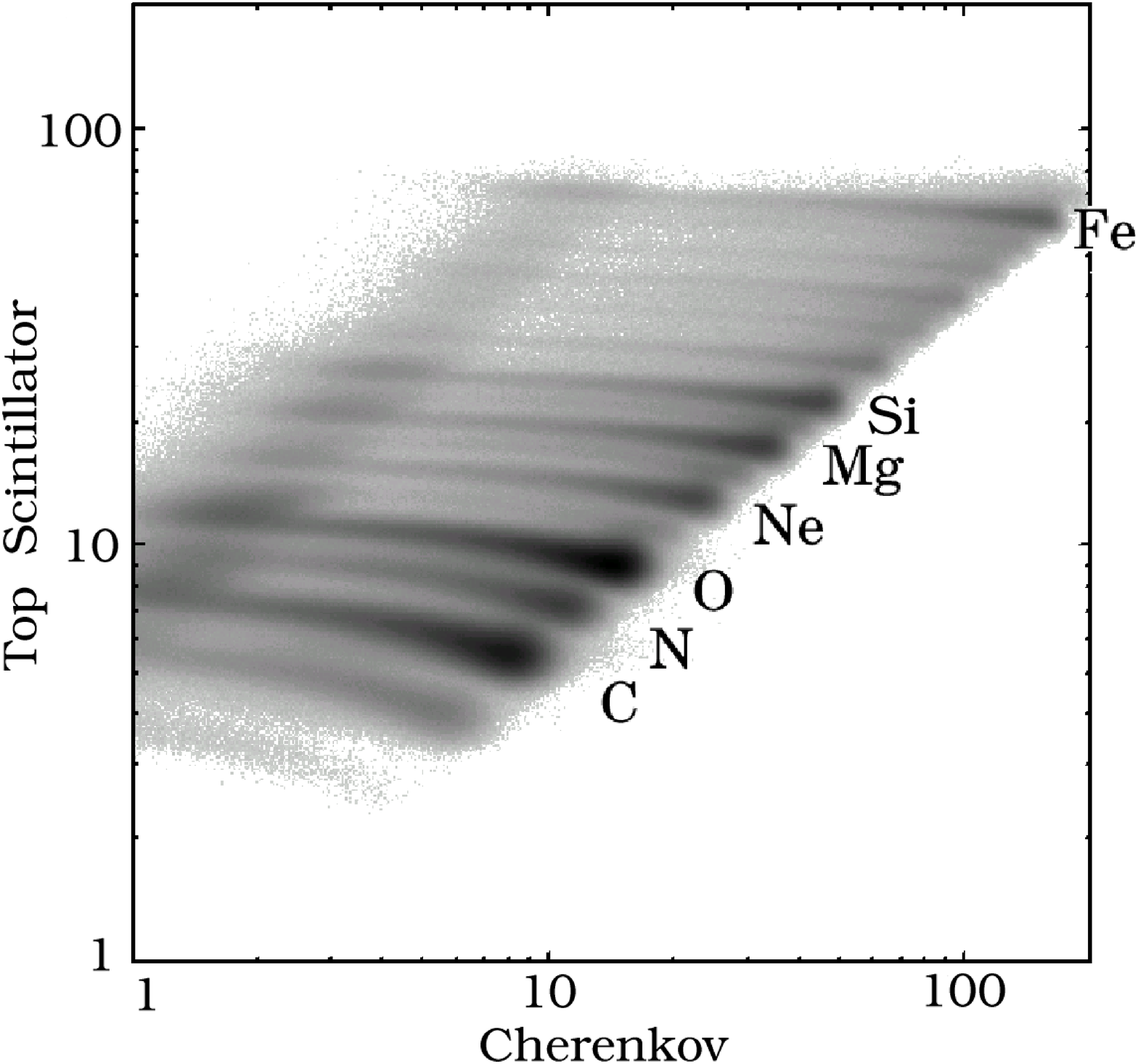}
\end{center}
\caption{Scatter plot of top scintillator vs. Cerenkov signal in arbitrary units.}
\label{CerScint}
\end{minipage}\hfill
\begin{minipage}[t]{.45\textwidth}
\begin{center}
\includegraphics[width=.99\textwidth]{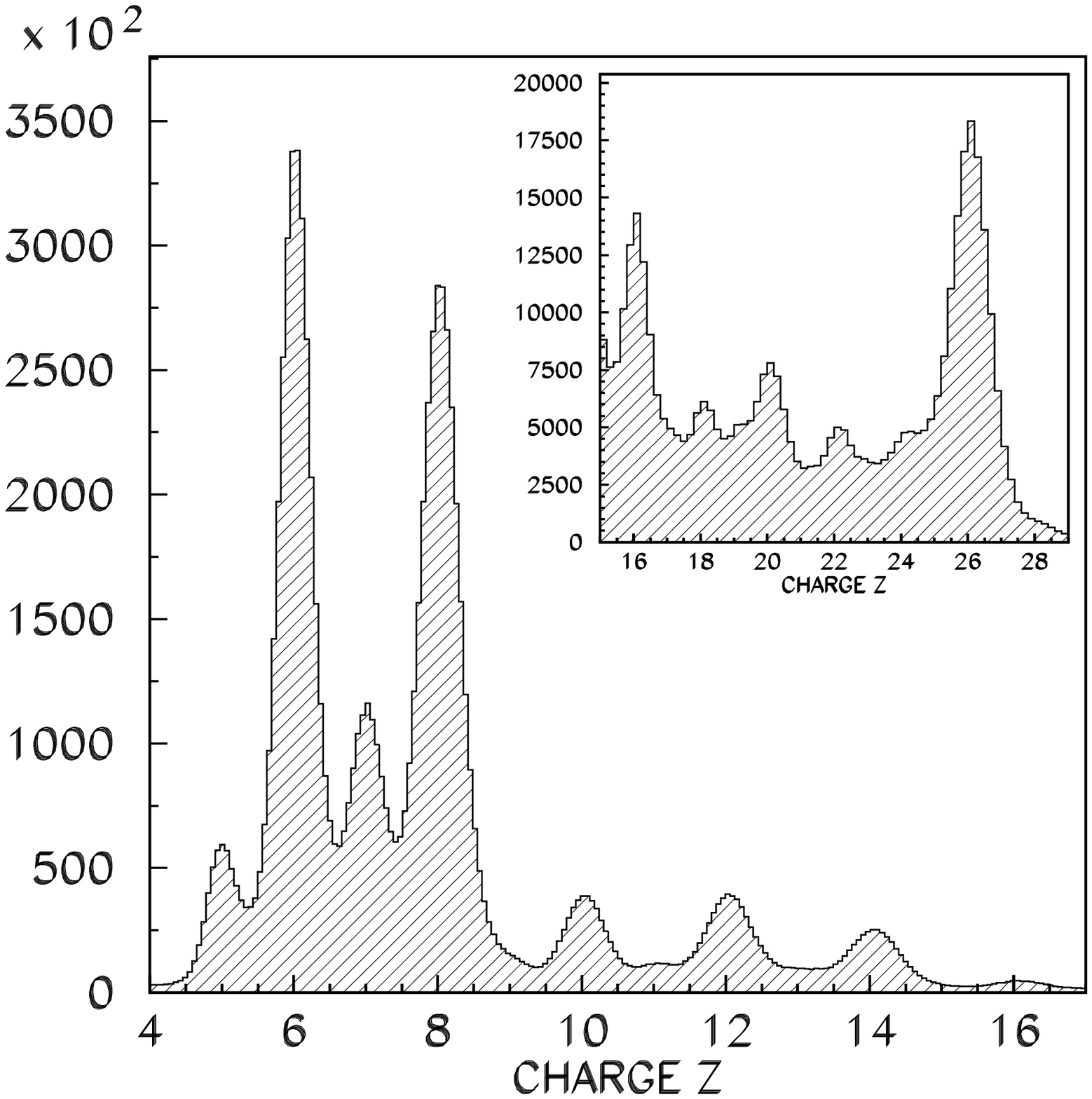}
\end{center}
\caption{Charge histogram for all events measured in flight.}
\label{Zhisto}
\end{minipage}
\end{figure}

The next stage in the analysis is to determine the charge of each
particle using a combination of the plastic scintillation and Cerenkov
counters. Trajectory information is used to correct for variations in
zenith angle. Spatial variations in the scintillation and Cerenkov
counters are corrected using both independent measurements with muons
on the ground and cosmic rays collected during the flight. By
combining the responses of scintillation and Cerenkov counters we
resolve individual charges (see figure \ref{CerScint}). Figure
\ref{Zhisto} shows a charge histogram for all charges obtained from
summing along lines of constant charge in figure \ref{CerScint}.


\begin{figure}[h]
\begin{minipage}[t]{.49\textwidth}
\begin{center}
\includegraphics[width=.99\textwidth]{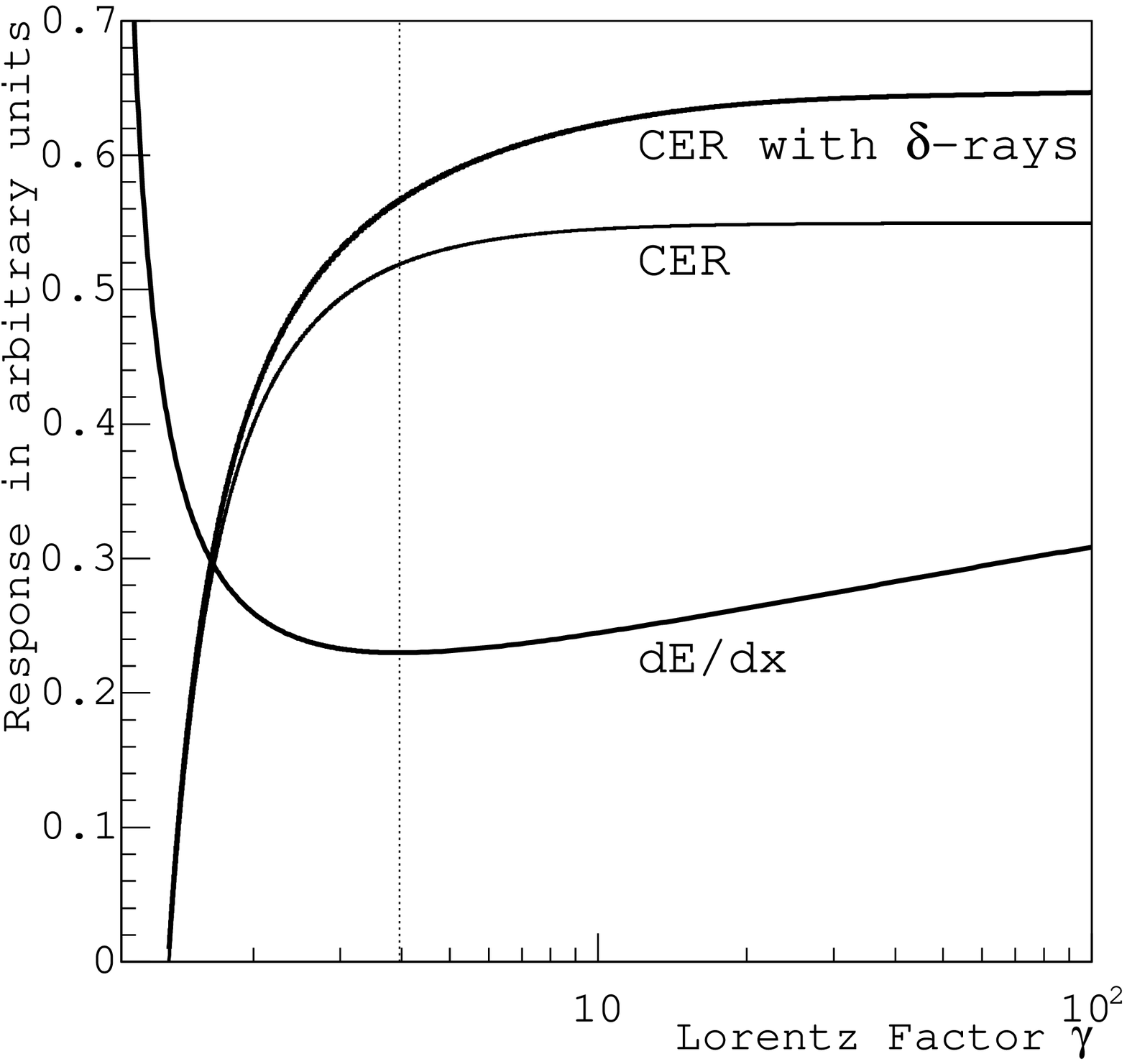}
\end{center}
\caption{Energy response of the Cerenkov counter with and without
taking into account the $\delta$ rays and response function of the
specific ionization detector. The dashed line indicates
minimum ionization.}
\label{CER}
\end{minipage}\hfill
\begin{minipage}[t]{.49\textwidth}
\begin{center}
\includegraphics[width=0.99\textwidth]{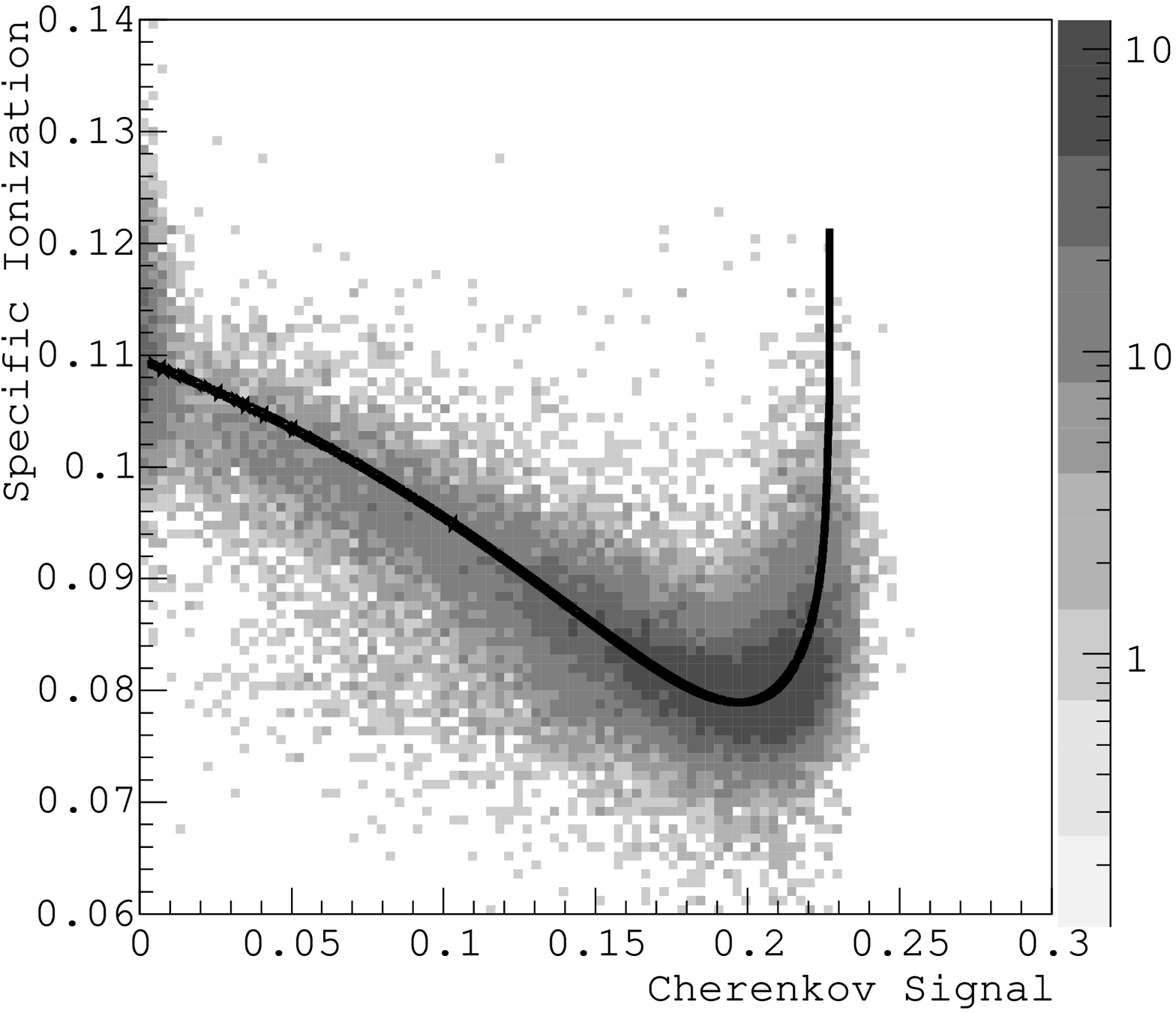}
\end{center}
\caption{Scatter plot of dE/dx vs. Cerenkov signals for iron
nuclei. The black line is the average response obtained from
simulations.}
\label{CerdEdx}
\end{minipage}
\end{figure}

\subsection{Identification of Low Energy Particles ($<$ 3 GeV/amu)}
\label{sec:lep}

A balloon flight near the geomagnetic pole exposes the instrument to a
large flux of low energy and sub relativistic particles which
overwhelms the flux in the TeV region by a factor of 10$^4$.  These
particles must be individually identified and separated with high
efficiency from the high energy data set. The separation is achieved
by combining the signal from the Cerenkov counter with that of the
ionization counter. The energy response for both counters are shown in
figure \ref{CER}. The ionization response, well described by the
Bethe-Bloch formula, is represented by the curve labeled dE/dx and the
Cerenkov response by the curve labeled CER. Note that the response of
the Cerenkov counter must be modified to account for effects by delta
rays generated by the primary particle in the instrument. This effect
is well understood and has been previously studied and reported
(\cite{gahbauer03b}). The signals from the proportional tube array
remain unaffected by delta rays (\cite{romero05b}). Figure
\ref{CerdEdx} illustrates the combined response of the ionization and
Cerenkov counters. The black line represents the averaged response
obtained from simulations. We see that the Cerenkov signal increases
while the ionization signal decreases towards minimum ionization
(CER=0.2, Ionization=0.08). At higher energy, the Cerenkov signal saturates
and the ionization signal increases due to the relativistic rise in
gases. Superimposed in gray are flight data from iron. The data
follow closely the pattern predicted by simulations. Although the
large flux of low energy particles represents a source of background
for the identification of particles, it also permits to determine
exactly the signal levels of minimum ionizing particles at 3
GeV/amu. The Cerenkov signals also are used to determine the energy
spectra of each particle species around 1 GeV/amu.


\begin{figure}[h]
\begin{minipage}[t]{.49\textwidth}
\begin{center}
\includegraphics[width=.99\textwidth]{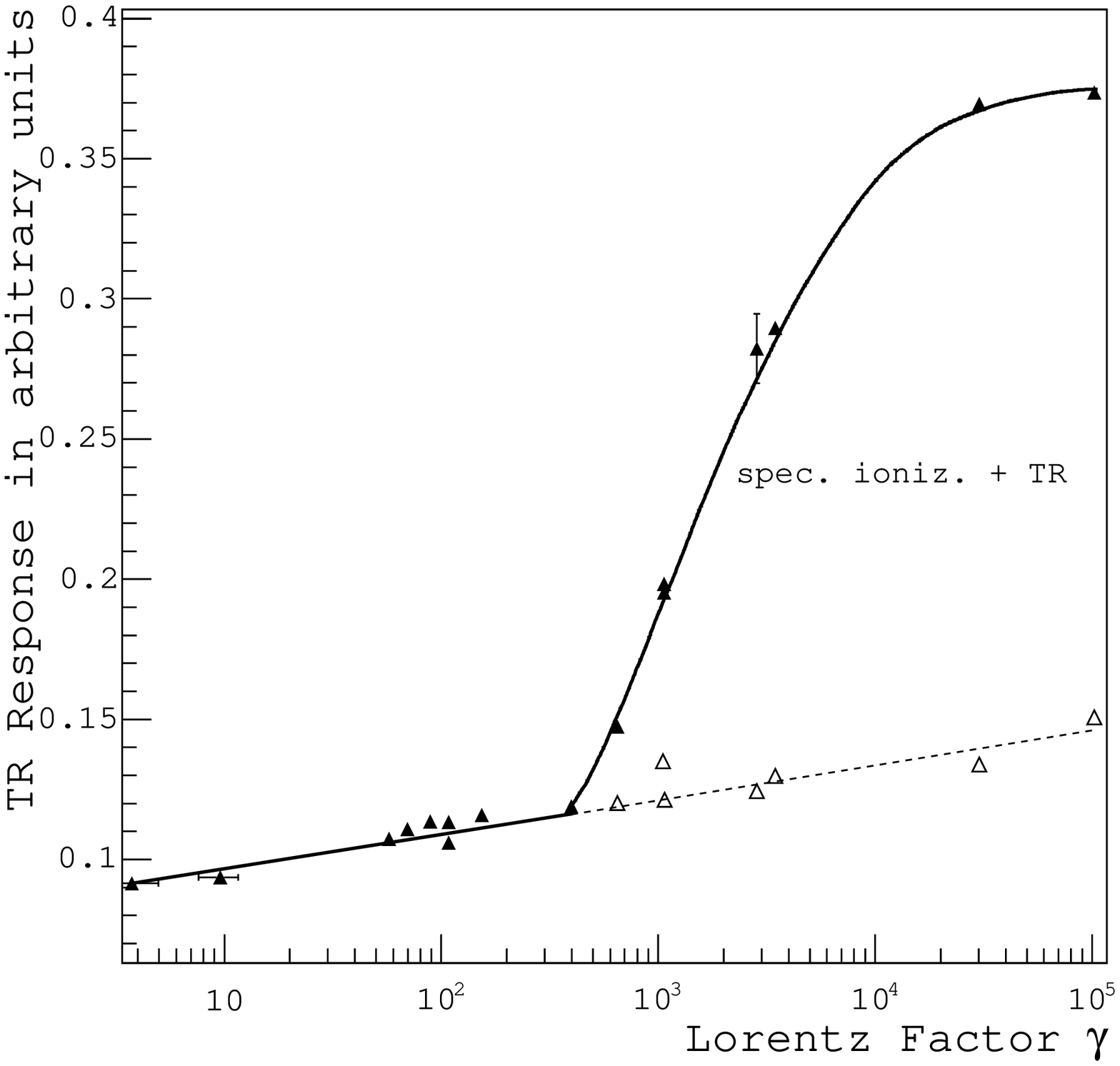}
\end{center}
\caption{Energy response of a Transition Radiation Detector used in
CRN, as measured at an accelerator \cite{lheureux90}}
\label{TRcurve}
\end{minipage}\hfill
\begin{minipage}[t]{.49\textwidth}
\begin{center}
\includegraphics[width=.99\textwidth]{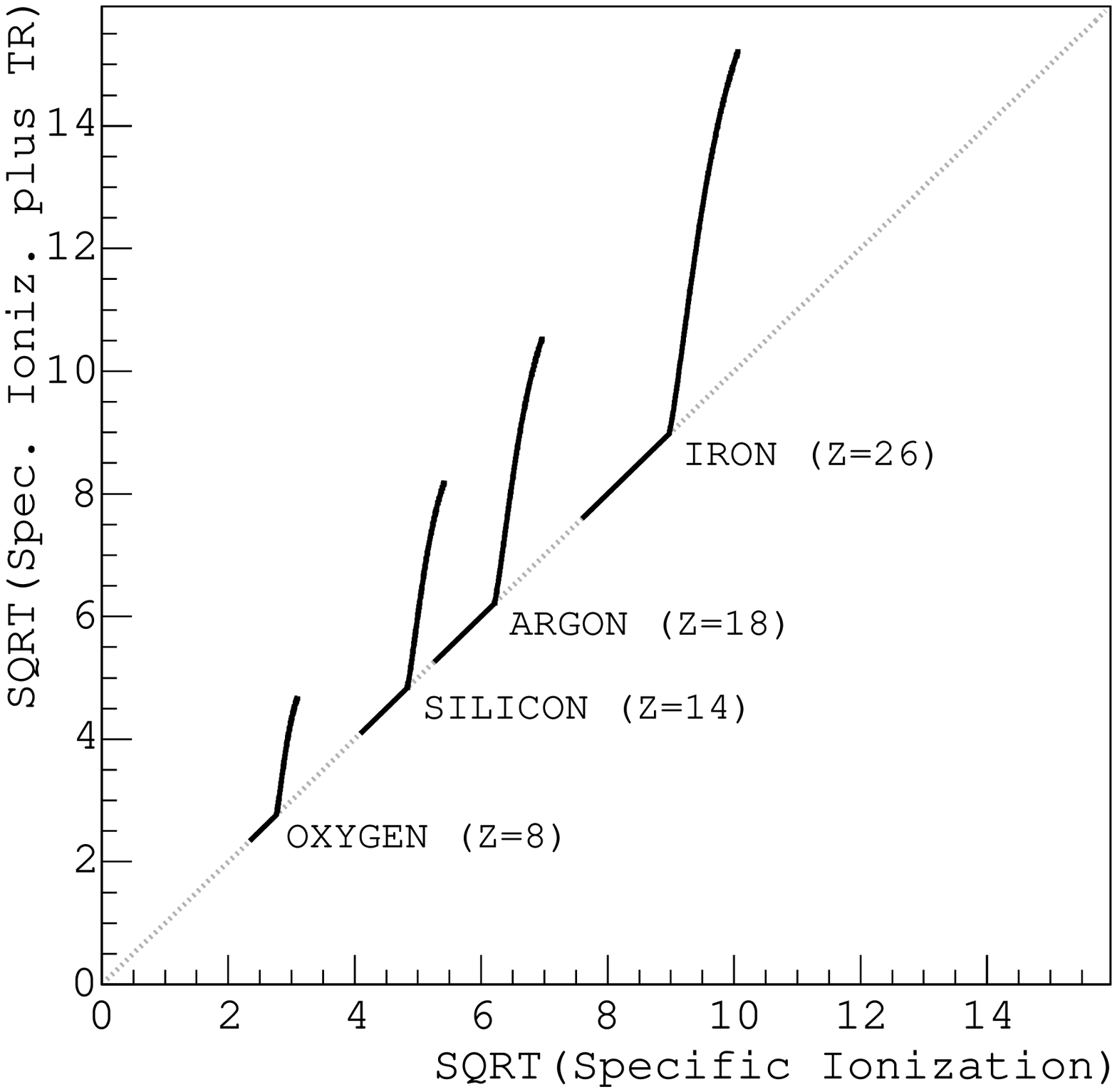}
\end{center}
\caption{Correlation of responses of Transition Radiation and specific
ionization detectors. Four elements are displayed to illustrate the
charge dependence of the responses.}
\label{TRDvsDEDX}
\end{minipage}
\end{figure}

\subsection{Identifying the Highest Energy Particles ($>$ 400 GeV/amu)}

After removing all particles below 3 GeV/amu using the method
detailed in section \ref{sec:lep}, the process of identifying the
rare high energy particles is achieved through a combination of the
measured responses of the ionization counter and the Transition
Radiation detector. One of the advantages of using these types of
detector is that they can be calibrated at an accelerator using singly
charged particles.  Figure \ref{TRcurve} shows the measured response
for the CRN instrument. Most of the data shown in this figure have
been published before (\cite{lheureux90}). For TRACER, the radiator
combination and gas mixtures in the detector are identical to that of
CRN, hence the CRN calibration remains valid. The dashed line is the
response of the ionization counter, which for TRACER consists of the
upper 800 proportional tubes and the solid line represents the TRD,
consisting of the lower 800 proportional tubes and radiator
combination. At energies between a few GeV/amu and 400 GeV/amu we
expect no observable transition radiation. Thus, both signals increase
logarithmically with energy, and on average, will lie along the
diagonal line in a correlation plot of TRD response vs ionization
response (see figure \ref{TRDvsDEDX}). Above 400 GeV/amu, TR becomes
observable and the signal from the TRD will increase above the
ionization signal. This manifests itself as a deviation from the
diagonal in the correlation plot. We expect the signals from
individual species to be well separated due to the Z$^2$ dependence of
the signal, with the fluctuations decreasing as 1/Z
(\cite{swordy90},\cite{wakely02}).

As an example, figure \ref{TRdEdx} shows the observed cross correlation
between the TRD and the ionization signals for neon nuclei (Z
= 10) above minimum ionization. The small black points represent the
numerous events with energies below the onset of TR. The rare high
energy particles with clear TR signals are highlighted. As expected,
the data follow the response illustrated in figure
\ref{TRDvsDEDX}. Note that the highest energy events (the \emph{TR
events}) stand out without any background in other regions of the
scatter plot. The most energetic neon nucleus in this sample of data
has a trial energy of $6 \times 10^{14}$eV.

\begin{figure}[h]
\begin{center}
\includegraphics[width=.5\textwidth]{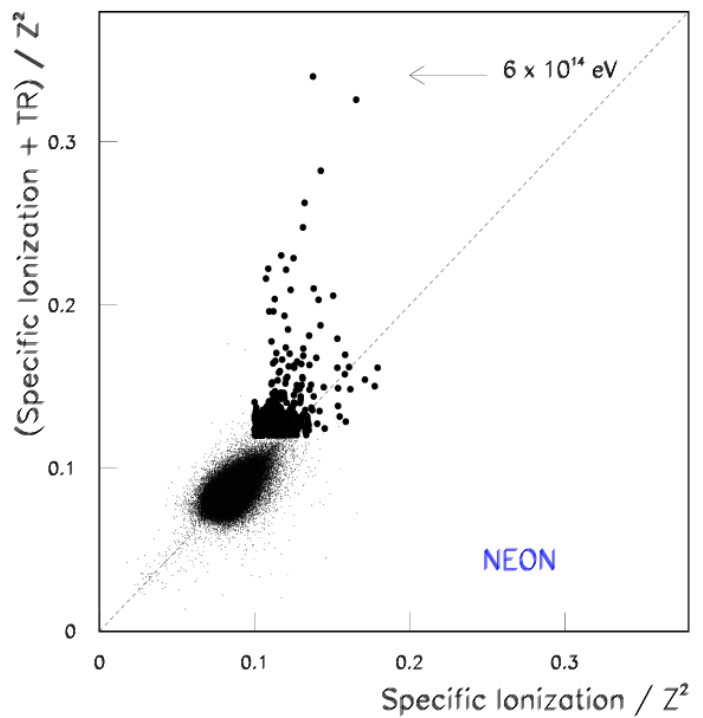}
\end{center}
\caption{Scatter plot of TR vs. dE/dx signal for neon nuclei. The
highlighted points represent the highest energy events measured with
the TRD. As expected the transition radiation events have signals in
the dE/dx detector which are well above the minimum ionization level.}
\label{TRdEdx}
\end{figure}

\subsection{Derivation of an Absolute Energy Spectrum}

We now derive absolute energy spectra over five decades of energy from
a few hundred MeV/amu to greater than a TeV/amu. We show here, as an
example, the spectrum for neon (Z = 10) at the top of the atmosphere.
This requires accurate knowledge of exposure factors and efficiencies
of the instrument. Table \ref{eff} lists the efficiencies due to cuts
made on the data which are divided into two data sets : (1) Low Energy
(below 3 GeV/amu) and (2) High Energy (above 3 GeV/amu).

The abundance of Low Energy data allows us to make a very strict cut
on the zenith angle, thus removing any angle effects in the Cerenkov
counter response. Energy bins have been chosen conservatively so that
overlap corrections are less than 20$\%$. The center of the energy
interval is calculated according to the method of
\cite{lafferty95}. The energy resolution for neon is 15\% at 1 GeV/amu,
80\% at 100 GeV/amu and 15\% at 1000 GeV/amu.

\begin{table}[h]
\begin{center}
\caption{Exposure and cut efficiencies, i.e. fractions of surviving particles for neon for the 2003 flight.}
\begin{tabular}{lcc}  
\bf{Cut/Efficiency} & \bf{Low Energy}  &  \bf{High Energy} \\ 
Zenith Angle & 29 - 30$^\circ$ & 0 - 60$^\circ$ \\
Tracking Efficiency & 0.95 & 0.95  \\
Top Scint Efficiency & 0.93 & 0.83  \\
Bot Scint Efficiency & 0.95 & 1.00 \\ 
\bf{Exposure} (m$^2$ str secs)& \bf{76044} & \bf{1664909}   \\
\end{tabular}
\label{eff}
\end{center}
\end{table}

\section{Discussion of Results}

\subsection{Comparison with previous Measurements}

\begin{figure}[h]
\begin{center}
\includegraphics[width=.5\textwidth]{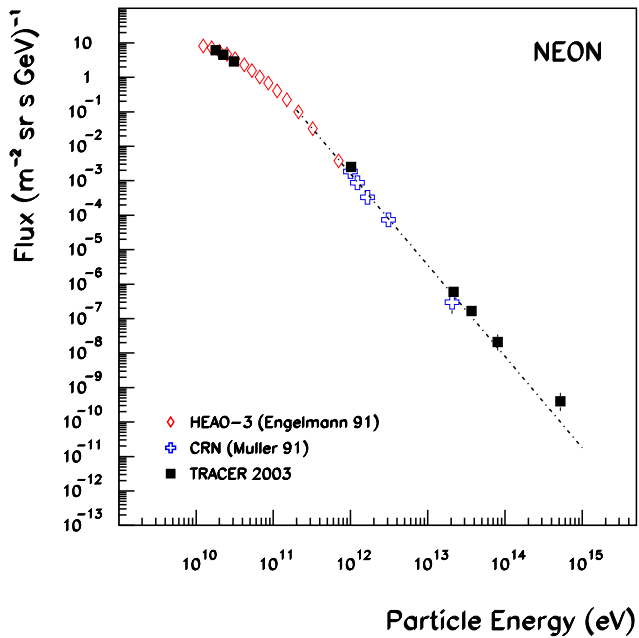}
\end{center}
\caption{Differential energy spectrum for neon from TRACER (solid
squares), HEAO-3 (open diamonds) and CRN (open crosses). The dashed
line represented a power law of $E^{-2.65}$.}
\label{spectrum}
\end{figure}

The absolute energy spectrum for the single element neon as measured
by the TRACER instrument from the 2003 flight is presented as black
squares in figure \ref{spectrum}. Again, we emphasize that all values
are given as absolute intensities with no arbitrary normalizations. We
first note the large range in terms of intensity (12 decades) and
particle energy (almost 5 decades) covered by TRACER. This has been
achieved by combining three complementary measurements in one detector
: the Cerenkov counter ($<10^{11}$eV), the relativistic rise of the
ionization signal in gas ($10^{11}-10^{13}$eV) and the TRD
($>10^{13}$eV)

TRACER has determined the individual energy spectrum for neon above
$10^{14}$eV for the first time. The spectrum does not indicate any
steepening towards $10^{15}$eV. Below $10^{12}$eV we compare our
results with measurements from the HEAO-3 satellite and at higher
energies with data from CRN on the Space Shuttle. Figure
\ref{spectrum} illustrates the good agreement in absolute intensity
between the TRACER data and the results from the space borne
detectors. The dashed line represents a power law of E$^{-2.65}$ and
describes the combined data set above $10^{11}$eV. A slight flattening
of the spectrum above $10^{14}$eV cannot be excluded, and its
significance will be discussed in the next section.

\subsection{Propagation in the Galaxy}

\begin{figure}[h]
\begin{center}
\includegraphics[width=.5\textwidth]{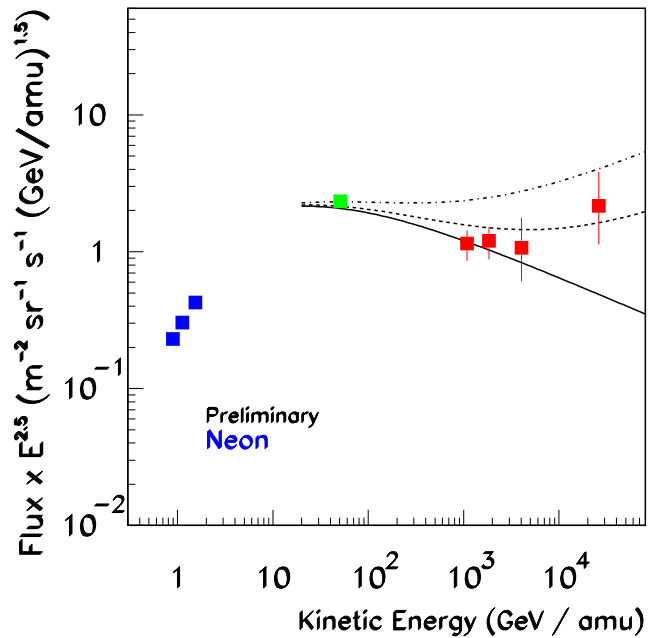}
\end{center}
\caption{Differential energy spectrum, multiplied by $E^{2.5}$, for
  neon as measured by TRACER. The curves refer to a prediction of the
  simple propagation model from the text. Solid line = $\Lambda_0$ = 0
  g/cm$^2$, dashed = 0.15 g/cm$^2$, dashdot = 0.5 g/cm$^2$}
\label{neon22}
\end{figure}

The results from HEAO-3 and CRN have previously been parameterized in
a simple propagation model which assumes a dynamic equilibrium between
particle production at the source and loss from the Galaxy by
diffusive escape or spallation. Good agreement with the measurements
has been found if the differential source spectra are proportional to
$E^{-2.2}$ and if the escape path length varies as $\Lambda \propto
R^{-0.6}$ for rigidities R $>$ 20 GV (\cite{swordy93}). Figure
\ref{neon22} (solid line) presents the TRACER spectrum for neon as
compared to a fit using this model. Note, that for clarity the flux
has been multiplied by $E^{2.5}$. The assumption of the model that the
escape path length continues as $E^{-0.6}$ up to arbitrary high
energies may not be realistic.

A more likely approach may be to introduce an energy independent
residual path length $\Lambda_0$ in addition to the escape path
length, i.e. $\Lambda(E) = AE^{-0.6} + \Lambda_0$. Figure \ref{neon22}
shows three scenarios for different values of the residual pathlength
: $\Lambda_0$ = 0 g/cm$^2$ (solid), $\Lambda_0$ = 0.15 g/cm$^2$
(dashed), and $\Lambda_0$ = 0.5 g/cm$^2$ (dashdot). As can be seen,
the value of $\Lambda_0$ cannot be much more than 0.15 g/cm$^2$ on the
basis of this simple analysis. However, determining such a limit with
more precision is still work in progress. We are extending the model
to the entire data set of the eight individual elements sampled by
TRACER and will soon report our conclusions. It would be highly
desirable to measure $\Lambda_0$ directly by determining the abundance
ratio between secondary and primary elements at high energies. An
ideal candidate is the boron to carbon ratio. However, the inclusion
of these light elements in the measurement was not possible to
accomplish in the 2003 flight due to a limited dynamic range in the
readout electronics of the TRD.

\section{Arctic Flight 2006}
\label{arctic}

We have concluded a second long duration balloon flight of TRACER in
2006, with a refurbished and upgraded instrument launched from Kiruna,
Sweden. The instrument was modified to permit a measurement of the
secondary to primary ratio, namely the ratio of boron to carbon.

\subsection{Upgrades on previous flights}

A measurement of the B/C ratio requires high precision in the charge
determination. Primary carbon is much more abundant than the secondary
boron, and a misidentified carbon nucleus could masquerade itself as
boron. To improve the charge resolution over the 2003 flight we
doubled the number of PMTs on the bottom scintillation counter, and
added a second plastic Cerenkov counter that was placed on top of the
instrument.  In the 2003 flight, the range of elements probed was
limited to Z=8 to 26 due to a limited dynamic range of the ASIC AMPLEX
chips (\cite{beuville90}) used to read out the signals from the
proportional tubes. To extend the range to include boron we split the
signal resistively for each proportional tube into a Low and High Gain
channel, thereby achieving an effective dynamic range of 12 bits or
4096 ADC channels. The doubling of the number of effective
proportional tube channels to 3200 required the use of a new DAQ
system, the heart of which is a custom Field Programmable Gate Array
system.

\subsection{Arctic Balloon Flight}
\begin{figure}[h]
\begin{center}
\includegraphics[width=.5\textwidth]{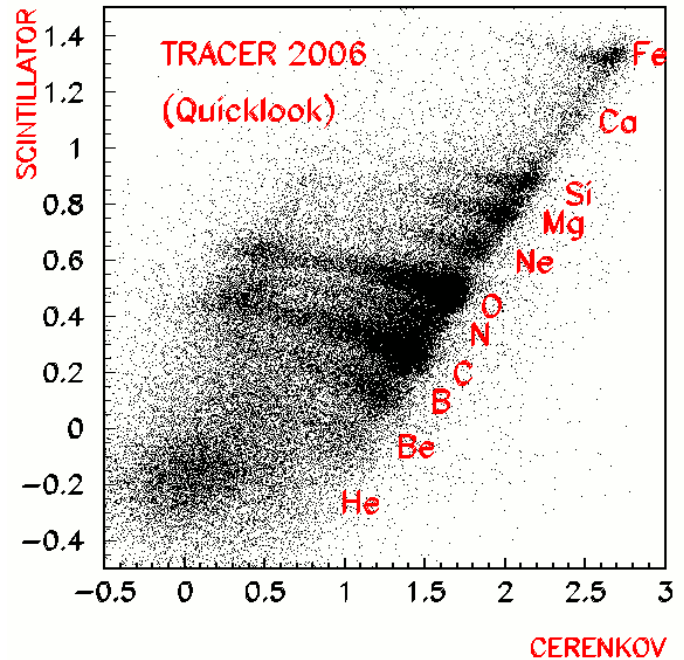}
\end{center}
\caption{Charge Distribution from TRACER 2006. These results are from
a crude \emph{Quicklook} analysis of 3 hours of data taken during Line
of Sight Telemetry. Units are arbitrary.}
\label{tracer2006}
\end{figure}

TRACER was launched from Kiruna, Sweden at 5am on July 8th 2006. The
balloon traveled westward over the Atlantic at an average altitude of
125,000 feet. The initial intent was to let the balloon circumnavigate
the North Pole and travel over Russia and land in Sweden or in
northern Canada. Unfortunately, Russian overflight permission was not
granted for this flight, and hence the flight had to be terminated
after four and a half days in northern Canada. The instrument was
successfully recovered and the onboard data disks were removed. Figure
\ref{tracer2006} shows a preliminary charge distribution for a sample
of data taken during this flight. The distribution demonstrates that
the TRACER instrument is sensitive to charge as low as Helium
(Z=2). The data analysis is presently underway.

\section{Conclusion}

The unique feature of the TRACER instrument is the use of gaseous
detectors (single wire proportional tubes) to measure both the
specific ionization, and specific ionization plus x-ray transition
radiation, in order to determine the energy of highly relativistic
nuclei (boron to iron). As compared to more conventional techniques,
this concept achieves a very favorable ratio of sensitive area to
weight; hence, the geometric factor of TRACER by far exceeds that of
all other current instruments. The technique also provides a high
degree of redundancy because of the multiplicity of measurements
taken, and it has the advantage that a complete calibration at
accelerators can be and has been performed.

TRACER has been flown in two long duration balloon flights, in 2003
and 2006. The data analysis of the 2003 flight is nearly complete and
demonstrates the power of the technique, determining the energy
spectra of individual cosmic ray nuclei ( 8 $\leq$ Z $\leq$ 26) over
nearly five orders of magnitude in intensity (\cite{boyle05}). The
small flux of particles at the highest energies is vastly out numbered
by cosmic rays of lower energy, but is identified cleanly and without
any contamination by low energy background. Therefore, the TRACER
results provide a sample of cosmic ray data, that with single charge
resolution, extends to the highest energies currently covered in
direct measurements, well in excess of $10^{14}$eV/particle.

While the results do not reveal any surprising features in the cosmic
ray energy spectra at high energies, they begin to provide stringent
constraints on the conventional models on galactic propagation. The
analysis of the TRACER results in the context of these models is
currently still in progress. The second balloon flight of TRACER
(2006) will provide very important additional detail, as it will
include a measurement of light secondary cosmic ray nuclei, in
particular, of the element boron (Z = 5).


\section{Acknowledgments}
This work has been supported by NASA grants NAG5-5305 and NN04WC08G
and the Aerospace Illinois Space Grant. We gratefully acknowledge the
services of NASA, the Columbia Scientific Balloon Facility, NSF
Antarctic Program and ESRANGE.

\bibliographystyle{elsart-harv_no_url}
\bibliography{mybibl.bib}


\end{document}